\documentclass[aps,prb,twocolumn,shortbibliography,superscriptaddress,article]{revtex4-1}
\usepackage{epsfig}
\usepackage{epstopdf}
\usepackage{amsmath}
\usepackage{amsfonts}
\usepackage{amssymb}
\usepackage{hyperref}
\usepackage{bm}
\usepackage{makecell}
\usepackage{rotating}
\usepackage{hyperref}
\usepackage{multirow}
\usepackage{graphicx}
\usepackage{float}

\usepackage{graphicx}% Include figure files
\usepackage{dcolumn}% Align table columns on the decimal point
\usepackage{bm}% bold math
\usepackage{color}

\usepackage{tikz,xcolor,hyperref}

\definecolor{lime}{HTML}{A6CE39}
\DeclareRobustCommand{\orcidicon}{%
	\begin{tikzpicture}
	\draw[lime, fill=lime] (0,0)
	circle [radius=0.16]
	node[white] {{\fontfamily{qag}\selectfont \tiny ID}};
	\draw[white, fill=white] (-0.0625,0.095)
	circle [radius=0.007];
	\end{tikzpicture}
	\hspace{-2mm}
}

\foreach \x in {A, ..., Z}{%
	\expandafter\xdef\csname orcid\x\endcsname{\noexpand\href{https://orcid.org/\csname orcidauthor\x\endcsname}{\noexpand\orcidicon}}
}
\begin{document}

%\title{Revisiting the topological properties of XMg$_{2}$Bi$_{2}$ (X = Ca, Sr, Ba, Yb): hybrid-functional calculations reveal topology only due to inward surface relaxation}

\title{Revisiting the topological properties of XMg$_{2}$Bi$_{2}$ (X = Ca, Sr, Ba, Yb and Eu)}

\author{Antoni Facca\orcidA} 
\affiliation{International Research Centre Magtop, Institute of Physics, Polish Academy of Sciences,
Aleja Lotnik\'ow 32/46, PL-02668 Warsaw, Poland}

\author{Xujia Gong\orcidB}
\affiliation{International Research Centre Magtop, Institute of Physics, Polish Academy of Sciences,
Aleja Lotnik\'ow 32/46, PL-02668 Warsaw, Poland}

\author{Amar Fakhredine\orcidD}
\affiliation{Institute of Physics, Polish Academy of Sciences, Aleja Lotnik\'ow 32/46, 02668 Warsaw, Poland}

\author{Carmine Autieri\orcidC}
\affiliation{International Research Centre Magtop, Institute of Physics, Polish Academy of Sciences,
Aleja Lotnik\'ow 32/46, PL-02668 Warsaw, Poland}

\author{Asiyeh Shokri\orcidE}
\email{ashokri@magtop.ifpan.edu.pl}
\affiliation{International Research Centre Magtop, Institute of Physics, Polish Academy of Sciences,
Aleja Lotnik\'ow 32/46, PL-02668 Warsaw, Poland}

\date{\today}
\begin{abstract} 
Density functional theory is known to underestimate band gaps in semiconductors and to overestimate inverted band gaps, frequently exaggerating the predicted size of the topological phase diagram of materials. Employing hybrid functionals and calculating the topological invariants, we revisit the topological properties of compounds crystallizing in the CaAl$_2$Si$_2$-type and demonstrate that the overestimation of the inverted band gaps is particularly pronounced in compounds with this crystal structure. Among these, the class of XMg$_2$Bi$_2$ materials (X = Ca, Sr, Ba, Yb, and Eu) is topologically trivial for all considered cations. Our calculations show that these materials are narrow-gap semiconductors with direct band gaps of 0.24-0.34~eV, slightly decreasing with increasing the atomic weight of the element X. We confirm this by applying uniaxial strain and hydrostatic pressure, confirming these results. We emphasize that the experimental observation of surface states alone is insufficient to establish nontrivial topology, as trivial semiconductors may host surface states without a surface Dirac point. Consequently, since these materials are intrinsically topologically trivial, any experimentally observed topological signatures should be attributed to extrinsic effects such as doping or surface reconstruction.   
Our results underscore the importance of accurately treating electronic correlations when assessing topological character, even in materials containing heavy elements with strong spin-orbit coupling, such as bismuth.
\end{abstract}


\maketitle

\section{Introduction}
% general on DFT
Density functional theory (DFT) within local and semi-local exchange–correlation approximations, such as the local density approximation or the generalized gradient approximation (GGA)\cite{Perdew96}, is well known to underestimate fundamental band gaps in semiconductors and insulators\cite{doi:10.1021/cr200107z,PhysRevLett.100.146401}. In narrow-gap insulators with strong spin–orbit coupling, this deficiency can be particularly severe and often leads to band inversion as a numerical artifact. As a consequence, GGA-based calculations tend to predict a band inversion in trivial narrow-gap semiconductors. This overestimation of band inversion directly affects the predicted topological classification. Several high-throughput and machine learning approaches have been proposed to search for new topological materials\cite{doi:10.1021/acsami.5c25994,ullah2026txlfusionhybridmachine}. However, in most of these methods, the lack of a proper treatment of electron correlation can lead to false-positive predictions of topological insulators.

In particular, within standard GGA calculations, well-known materials such as PbTe\cite{Svane2010}, InAs\cite{Hussain2022electronic} and Ge\cite{Riemelmoser2026} are incorrectly predicted to exhibit inverted band ordering and are therefore classified as topological insulators or metals. However, it is well established from both experiments and more accurate theoretical treatments that these compounds are, in fact, topologically trivial narrow-gap semiconductors with normal band ordering. More advanced approaches, such as many-body perturbation theory within the $GW$ approximation, hybrid functionals, or the modified Becke–Johnson (mBJ) potential, significantly improve the description of band gaps and band ordering. When these methods are employed, the spurious band inversion found in GGA is removed, and the correct trivial topological nature of PbTe\cite{Svane2010}, InAs\cite{Hussain2022electronic}  and Ge\cite{Riemelmoser2026} is recovered. This highlights the necessity of going beyond semi-local DFT when assessing the topological properties of narrow-gap semiconductors. The Coulomb repulsion is effective on open shells such as d- and f-bands and is important for predicting the band gap and topological nature when d- or f-bands are at the Fermi level and involved in band inversion. The Coulomb repulsion U and other approaches to increase electronic correlations always make the compounds more trivial when there is one band inversion. In the case of more band inversions across the Brillouin zone, a correlation-induced topological phase transition is instead possible\cite{Hussain2023,D3CP01368E,PhysRevB.109.075147}.

Dirac semimetal phases can emerge with or without band inversion, depending on the underlying crystal symmetry and electronic band structure. In systems exhibiting band inversion, the reversal of conduction and valence band characters can facilitate the formation of symmetry-protected Dirac points\cite{Yang2014}. However, band inversion is not a prerequisite for realizing a Dirac semimetal phase, as Dirac points can also be stabilized by crystalline symmetries in systems with a conventional band ordering, as in symmetry-enforced Dirac semimetals\cite{PhysRevLett.108.140405}.  Consequently, Dirac semimetals can be broadly categorized based on whether their Dirac points arise from inverted or non-inverted band structures, reflecting the diverse mechanisms responsible for the emergence of Dirac fermions in quantum materials.\cite{RevModPhys.90.015001,wadge2026contemporaryinsightselectronicstructure}

% zero-gap topological insulators
%Unlike conventional topological insulators, bulk HgTe is a prototypical example of a zero-gap topological insulator\cite{PhysRevB.103.115209,PhysRevResearch.4.023114}. The absence of a band gap originates from the fourfold degeneracy of the $\Gamma_8$ valence-band states at the Brillouin-zone center. In the zinc-blende crystal structure, strong spin--orbit coupling inverts the normal band ordering, placing the $\Gamma_6$ conduction band below the $\Gamma_8$ states. Because the heavy-hole and light-hole components of the $\Gamma_8$ manifold remain fourfold degenerate at the $\Gamma$ point, the conduction and valence bands touch, resulting in a vanishing bulk band gap.
%A finite topological band gap can be induced by lifting the $\Gamma_8$ degeneracy through epitaxial strain, quantum confinement, breaking of inversion symmetry\cite{Islam2023,PhysRevB.110.165112,PhysRevB.104.L220404} or reduced crystal symmetry, thereby transforming HgTe into a three-dimensional or two-dimensional topological insulator, depending on the system geometry. The zero-gap topological insulator with preserved time-reversal symmetry is characterized by the topological invariant $\mathbb{Z}_2$\cite{PhysRevLett.95.146802}, with $\mathbb{Z}_2$=1 for bulk\cite{PhysRevB.76.045302} and quantum wells of HgTe\cite{Bernevig:2006_S,PhysRevB.107.045138}.

\begin{figure*}
    \centering
    \includegraphics[width=1\linewidth]{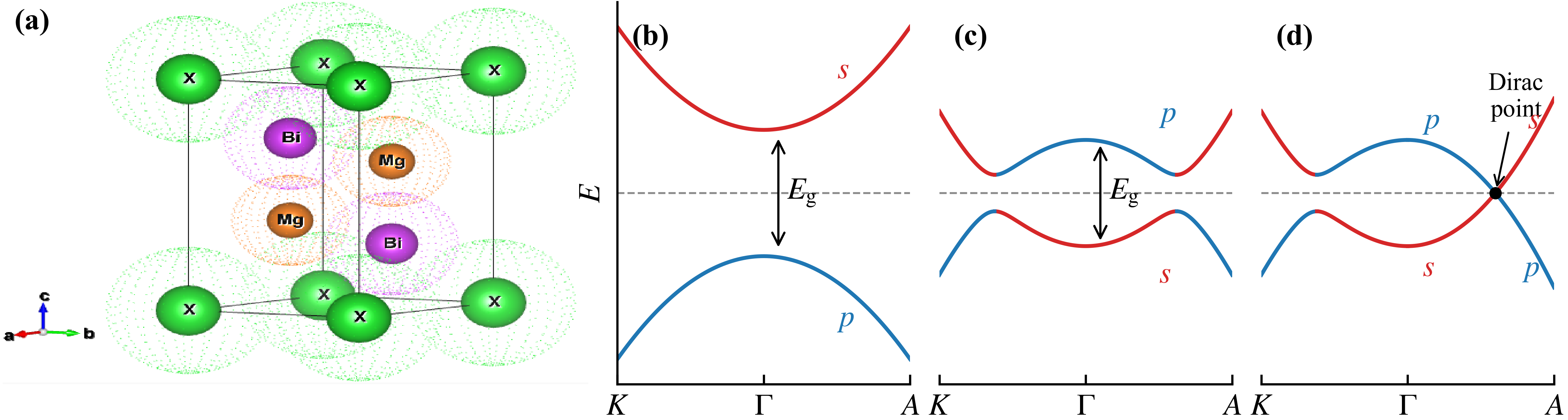}
    \caption{ (a) Crystal structure of XMg$_{2}$Bi$_{2}$ (X = Ca, Sr, Ba, Yb) in the CaAl$_2$Si$_2$-type structure with $P\overline{3}m1$ symmetry. (b) Schematic band structure of a trivial insulator with $\mathbb{Z}_2 = 0$ and $\mathbb{Z}_4 = 0$ with a direct band gap at $\Gamma$. (c) Schematic band structure of a topological insulator with $\mathbb{Z}_2 = 1$ and band inversion at $\Gamma$. (d) Schematic band structure of a three-dimensional Dirac semimetal with band inversion at $\Gamma$, characterized by $\mathbb{Z}_2 = 1$ and $\mathbb{Z}_4 = 3$. The Fermi level is set to zero and represented by a dashed line.}
    \label{fig:Figure1}
\end{figure*}

% particular case of Eu-based
In recent years, Eu-based compounds have emerged as a widely studied platform in this field. However, the narrow bandwidth of the Eu-$4f$ states generally makes them unsuitable for hosting topological bands, with EuIn$_2$As$_2$ and EuSn$_2$As$_2$ being notable exceptions~\cite{Cuono23EuCd2As2}. We have previously shown that DFT strongly overestimates the inverted band gap, and thus the topological character, in Eu-based systems crystallizing in the CaAl$_2$Si$_2$-type and related structures~\cite{Cuono23EuCd2As2}. Our analysis focused on EuCd$_2$As$_2$, which crystallizes in the CaAl$_2$Si$_2$-type structure with P$\overline{3}$m1 symmetry (space group No.~164)~\cite{Cuono23EuCd2As2}. In this compound, the band inversion involves Cd-$5s$ and As-$4p$ states, while Eu-derived bands do not participate. Consequently, applying a Hubbard $U$ to the Eu $4f$ states corrects their energies but does not remove the spurious inversion, thereby requiring more advanced approaches. We demonstrated that hybrid exchange-correlation functionals, particularly HSE, reliably correct the inverted gap and the resulting topological misclassification. These conclusions were independently confirmed by experiments showing a trivial band gap in EuCd$_2$As$_2$~\cite{PhysRevLett.131.186704}. The reliability of hybrid functionals was further validated by the HSE prediction of trivial electronic behavior in EuCd$_2$Sb$_2$~\cite{Cuono23EuCd2As2}, later verified experimentally~\cite{Sar2026-ob}.  

% we start to talk more about CaAl$_2$Si$_2$-type 
Focusing on materials crystallizing in P$\overline{3}$m1 symmetry with the CaAl$_2$Si$_2$-type structure and chemical composition XMg$_2$Bi$_2$ (X=Ca, Sr, Ba, Yb and Eu)\cite{Petrov2017,RAMIREZ2015217,doi:10.1021/ic2016808,PhysRevB.85.035202,doi:10.1021/acs.jpclett.5c00916,10.1039/c6cp02057g} as depicted in Fig. \ref{fig:Figure1}(a). 
In XMg$_2$Bi$_2$, the formal valence states are taken as X$^{2+}$, Mg$^{2+}$, and Bi$^{3-}$, consistent with the charge neutrality condition.
For the Mg$_3$Bi$_2$ compound, hybrid functionals predict a topological behavior with $\mathbb{Z}_2$=1 and a type-II nodal-line semimetal\cite{https://doi.org/10.1002/advs.201800897,doi:10.1021/acs.jpclett.7b02129}.
For XMg$_2$Bi$_2$ (X=Ca, Sr, Ba and Yb), it was claimed that CaMg$_2$Bi$_2$ and YbMg$_2$Bi$_2$ are putative topological insulators with $\mathbb{Z}_2$=1\cite{Kundu2022}, while it was claimed that BaMg$_2$Bi$_2$ is a Dirac semimetal with the Dirac point along the $\Gamma$-A direction\cite{Takane2021,ps93-rfk3}. However, these claims on XMg$_{2}$Bi$_{2}$ (X = Ca, Sr, Ba, Yb) are based on standard DFT band structure calculations, whereas experimentally the samples are p-type and the Dirac points in the topological surface states are inaccessible. 
In EuCd$_2$As$_2$, we have shown\cite{Cuono23EuCd2As2} that the effect of the Coulomb repulsion of Eu on the inverted band gap is minimal; therefore, we expect that the effect of the first cation (X in this paper) should not affect the band inversion in the first approximation, which depends mostly on Mg and Bi.

The three topological phases discussed in the literature are reported in Fig. ~\ref{fig:Figure1}(b,c,d), which are the trivial insulating phase in Fig. ~\ref{fig:Figure1}(b), the topological insulators in Fig.~\ref{fig:Figure1}(c) and the three-dimensional Dirac semimetals in Fig.~\ref{fig:Figure1}(d). In the latter case, two Dirac points are located at the position (0,0,$\pm{k^*_z}$) with ${k^*_z}\approx$~0.

% In this paper, we do
In this paper, we show that the members of the family XMg$_2$Bi$_2$(X = Ca, Sr, Ba, Yb and Eu) are, in fact, trivial semiconductors, exhibiting a direct band gap of  0.25-0.35~eV. These materials are intrinsically topologically trivial, and any realization of nontrivial topology must arise from extrinsic effects such as stoichiometric variations, crystallographic defects, doping, or surface reconstructions. By applying uniaxial strain and hydrostatic pressure, we further highlight that these compounds are robust trivial insulators.
The paper is organized as follows: In Section II, we report the electronic properties obtained using standard DFT and hybrid functional calculations, together with the evaluation of the topological invariants from ab initio results. In Section III, we apply uniaxial strain and hydrostatic pressure to investigate the robustness of the electronic phases. Finally, in Section IV, we draw our conclusions.

\section{Band structure and topological invariants}

In this Section, we report the band structure of XMg$_{2}$Bi$_{2}$ (X = Ca, Sr, Ba, Yb) within standard GGA calculations with the computation of their topological invariants. In the second subsection, we report the results using the hybrid functionals. Finally, we report a comparison with the Eu-based compounds. In all cases, the compounds exhibit Kramers degeneracy.

\subsection{Electronic properties of non-magnetic XMg$_2$Bi$_2$ with standard DFT}

\begin{figure}
    \centering
    \includegraphics[width=1\linewidth]{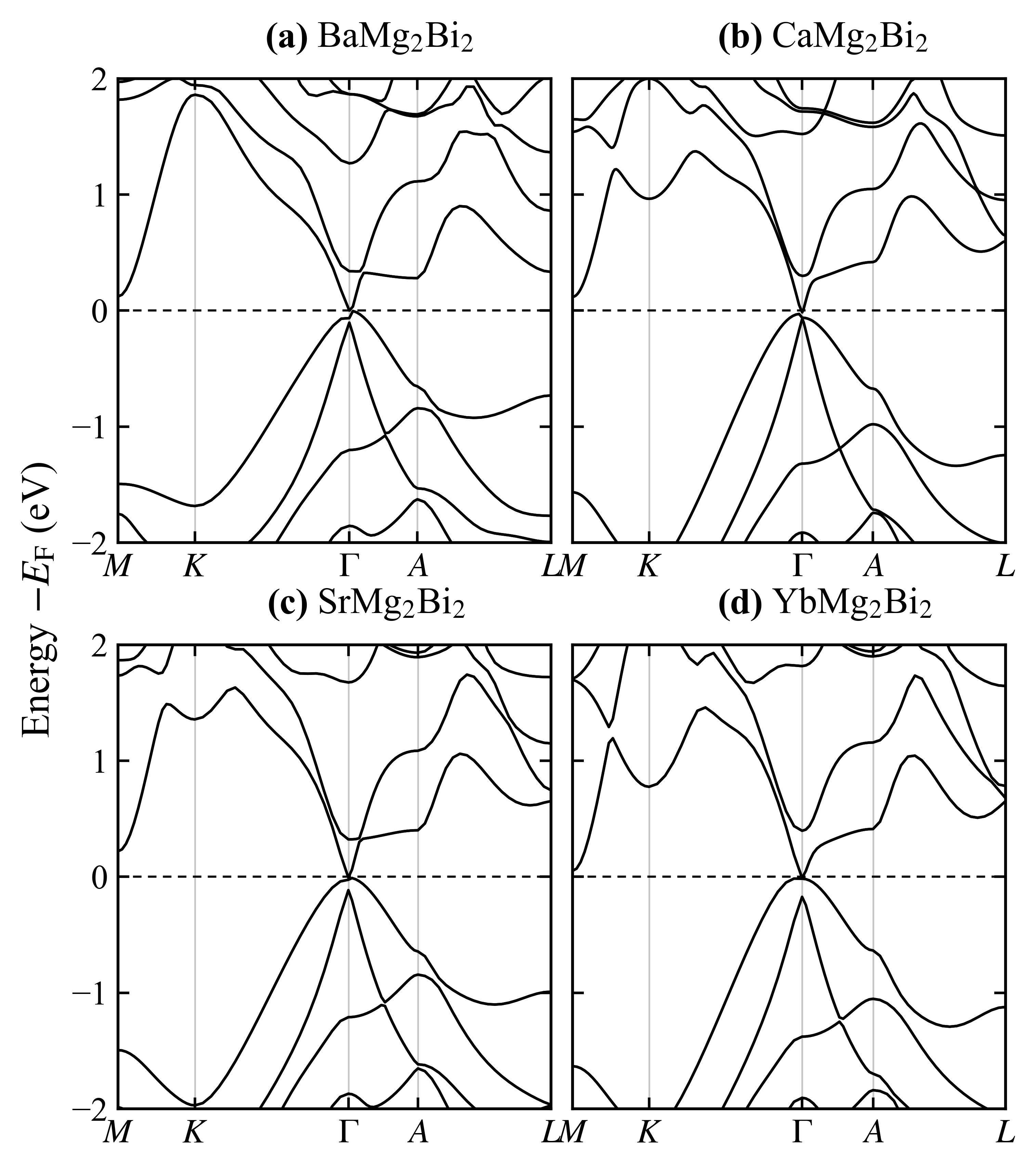}
    \caption{Relativistic band structures within GGA of XMg$_2$Bi$_2$ along the high-symmetry path M--K--$\Gamma$--A--L for (a) X = Ba, (b) X = Ca, (c) X=Sr, and (d) X = Yb. The energy range spans from $-2$ to $+2$~eV, with the Fermi level set to zero, indicated by a dashed line.}
    \label{fig:Figure2}
\end{figure}

The GGA band structure is reported in Fig. \ref{fig:Figure2}(a,b,c,d) for XMg$_{2}$Bi$_{2}$ (X = Ca, Sr, Ba, Yb). The band structures within GGA are qualitatively the same for all four compounds.
These phases are three-dimensional Dirac semimetals with the Dirac point along the $\Gamma$-A direction produced by the band inversion at the $\Gamma$ point.
For this kind of band structure, we also calculated the topological invariant and, differently from the literature, we find that these topological invariants are not trivial.

For all our XMg$_2$Bi$_2$ systems (X = Ca, Sr, Ba, Yb), we calculate the topological invariants within GGA and obtain trivial weak topological indices, $\mathbb{Z}_{2w,1}=\mathbb{Z}_{2w,2}=\mathbb{Z}_{2w,3}=0$, together with a nontrivial symmetry indicator, $\mathbb{Z}_4=3$. A similar behavior has been reported for $\beta$-As$_2$Te$_3$\cite{PhysRevB.104.024103}, which exhibits a three-dimensional Dirac semimetal phase characterized by the same topological indices. Also in that material, the Dirac semimetal state originates from a band inversion between the valence and conduction bands, giving rise to symmetry-protected Dirac crossings. It is important to note that the symmetry indicator value $\mathbb{Z}_4=3$ also implies a nontrivial strong topological index, $\mathbb{Z}_2=1$, indicating an underlying strong topological insulator phase.

\subsection{Electronic properties of non-magnetic XMg$_2$Bi$_2$ with hybrid functional}

\begin{figure}
    \centering    \includegraphics[width=1\linewidth]{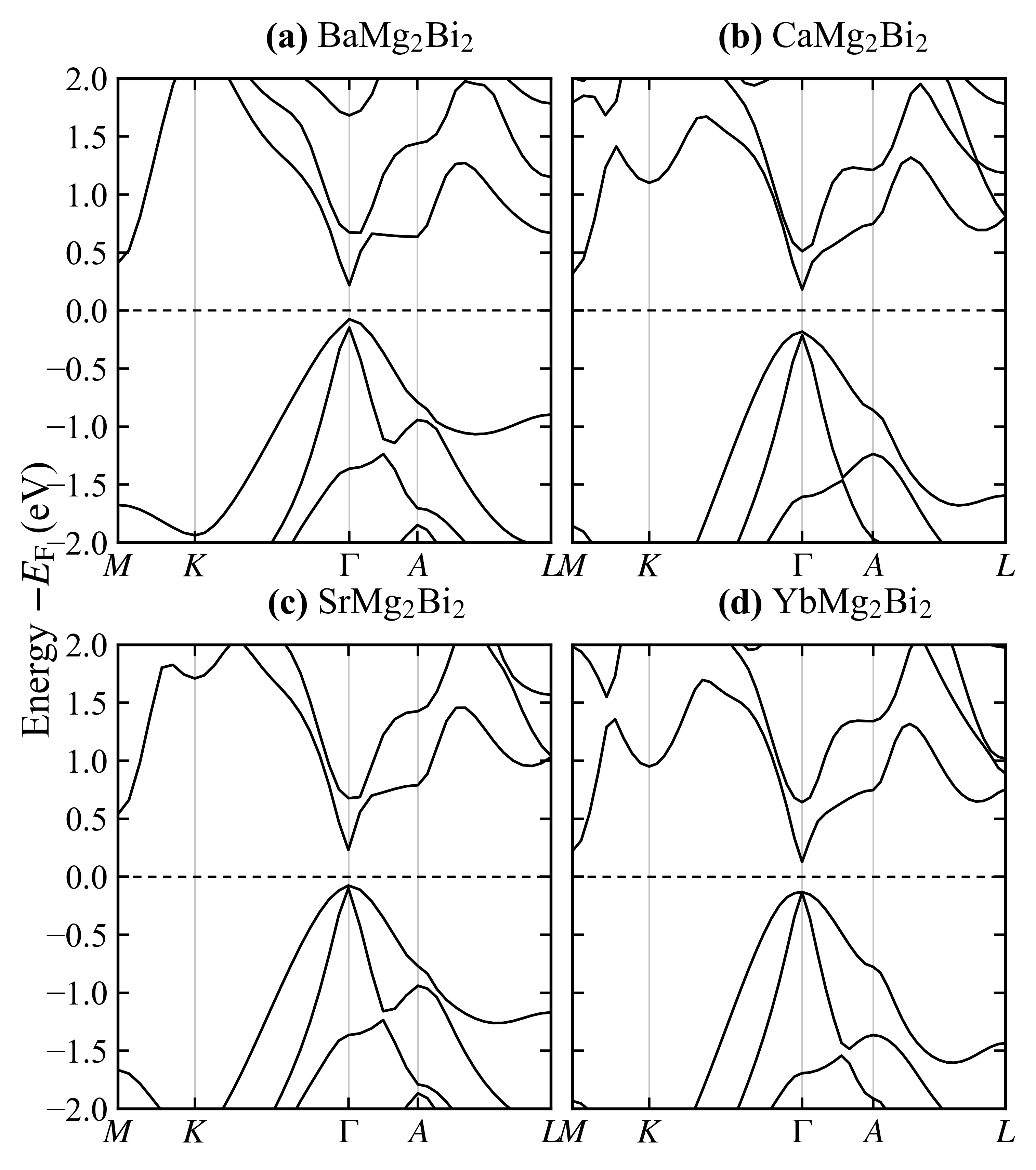}
    \caption{Relativistic band structures within HSE of XMg$_2$Bi$_2$ along the high-symmetry path M--K--$\Gamma$--A--L for (a) X = Ba, (b) X = Ca, (c) X=Sr, and (d) X = Yb. The energy range spans from $-2$ to $+2$~eV, with the Fermi level set to zero, indicated by a dashed line. The Fermi level is set on the top of the valence band.}
    \label{fig:Figure3}
\end{figure}

Here, we report our results using hybrid functionals for XMg$_2$Bi$_2$.
The relativistic band structure within hybrid functionals is reported in Fig. \ref{fig:Figure3}(a,b,c,d). The band structures with hybrid functionals are qualitatively the same for all four compounds.
The hybrid functional incorporates stronger electron correlation effects and opens the trivial band gap. In this case, applying GGA+U\cite{liechtenstein1995density} would not be appropriate, because the Hubbard correction primarily accounts for on-site Coulomb interactions in systems containing partially filled localized orbitals, such as 3d and 4f electrons. In this case, the shell of 3s-electrons of Mg is empty, while the shell of 6p-electrons of Bi is full. 
From the fat bands, we can observe for all compounds that the conduction band minimum is mainly derived from the 3$s$-states of Mg, while the valence band maximum originates primarily from the 6$p$-states of Bi. The compounds exhibit a direct band gap at the $\Gamma$ point.

For all our systems XMg$_{2}$Bi$_{2}$ (X = Ca, Sr, Ba, Yb), we calculate the topological invariants within hybrid functionals, obtaining $\mathbb{Z}_2=0$ and $\mathbb{Z}_4=0$. Therefore, within hybrid functional these compounds exhibit a trivial insulating phase.

\begin{figure}    
    \includegraphics[width=1\linewidth]{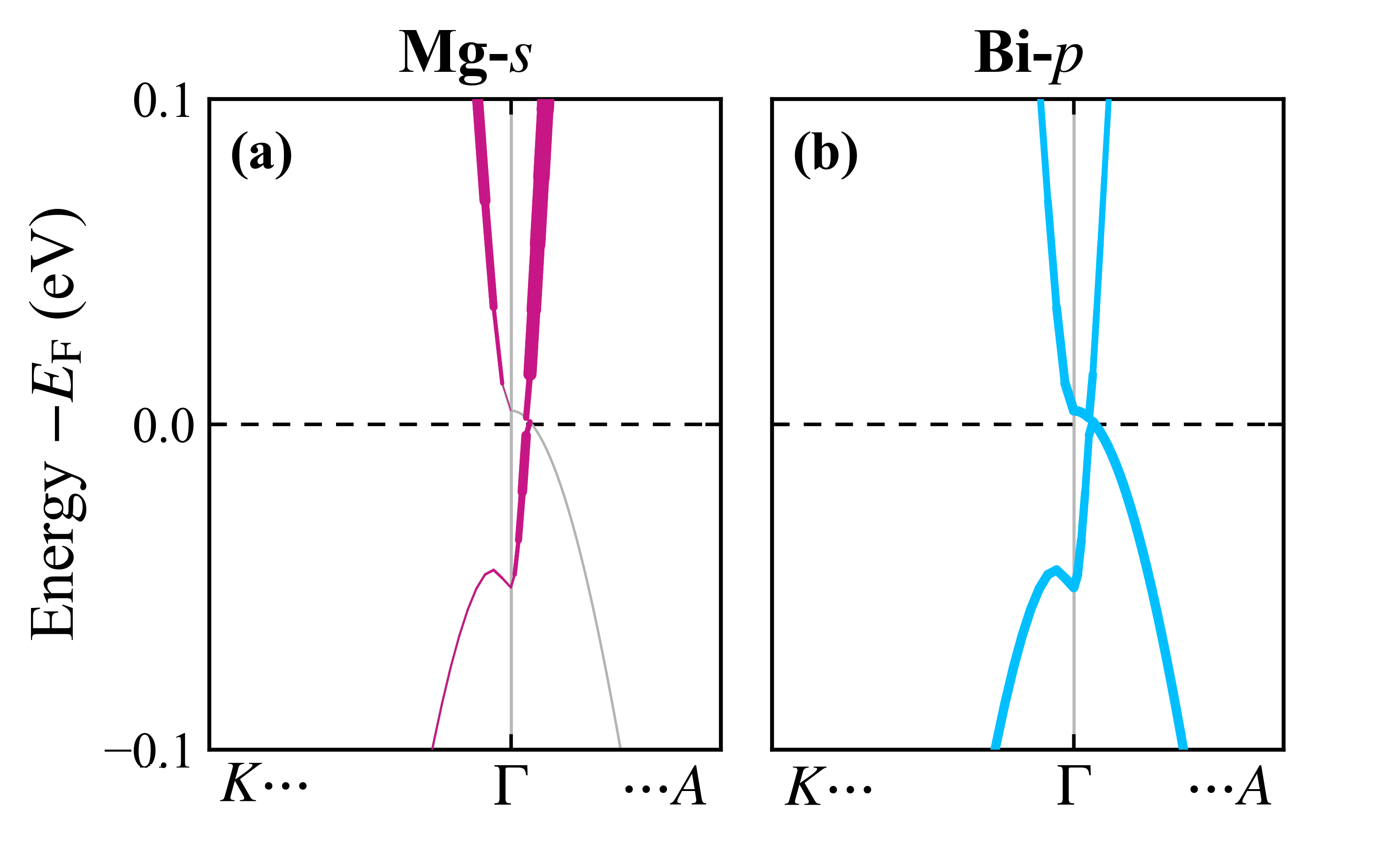}
    \includegraphics[width=1\linewidth]{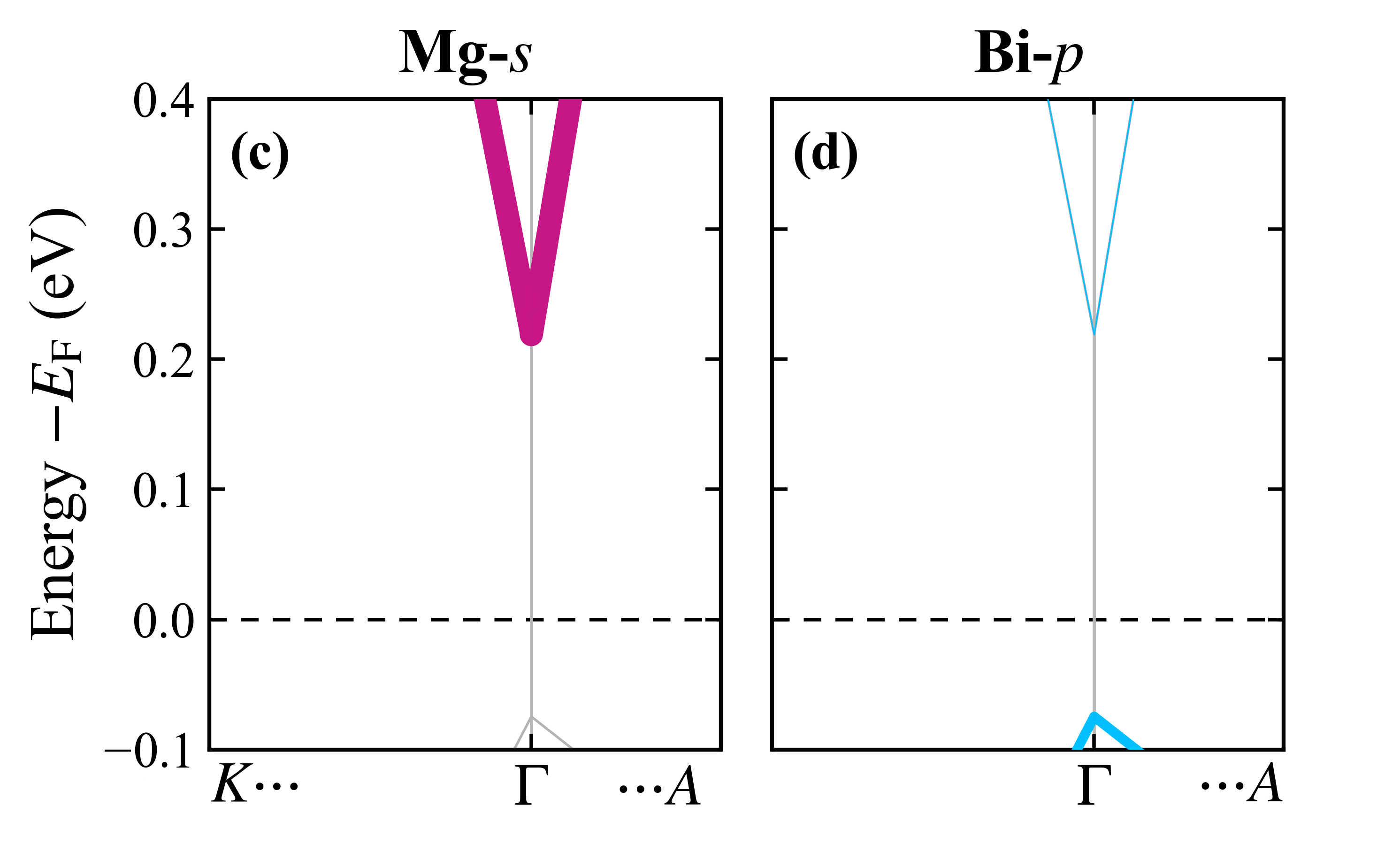}
    \caption{(a,b) Relativistic band structure of BaMg$_2$Bi$_2$ within GGA in the energy range between -0.1 and +0.1 eV. (c,d) Same as in (a,b) but within HSE between -0.1 and +0.4 eV. There is a clear separation between the Mg-s in the conduction band and the Bi-p in the valence band. The Fermi level is set to zero and represented by a dashed line. }
    \label{fig:Figure4}
\end{figure}
To highlight the differences between the GGA and hybrid functionals, we plot the fat-band representations of the Mg $3s$ and Bi $6p$ orbitals. Figures~\ref{fig:Figure4}(a,b) show the GGA calculations, whereas Figures~\ref{fig:Figure4}(c,d) present the hybrid-functional results. In the topological phase, the valence band is predominantly derived from the Mg $3s$ states, while the conduction band is mainly composed of the Bi $6p$ states. In the trivial phase, the orbital character is reversed. The GGA calculations predict the topological band ordering schematically illustrated in Fig.~\ref{fig:Figure1}(d), whereas the hybrid-functional calculations recover the trivial band ordering shown schematically in Fig.~\ref{fig:Figure1}(a).

\subsection{Comparison with Eu-based materials with CaAl$_2$Si$_2$-type structure}

Among magnetic compounds with the same crystal structure, there are EuCd$_2$As$_2$, EuCd$_2$Bi$_2$ and EuMg$_2$Bi$_2$. These compounds exhibit antiferromagnetic behaviour with Kramers degeneracy preserving time-reversal symmetry.

Previous studies have shown that EuCd$_2$As$_2$ within the GGA framework exhibits band inversion at $\Gamma$ with a Dirac point located along the $\Gamma$-A direction \cite{Ma:2020_AM} analogue to the results reported in subsection IIA. Within GGA, the calculated topological invariant $\mathbb{Z}_4=2$ indicates a distinct topological character corresponding to an axion insulator phase. This behavior differs from that of XMg$_{2}$Bi$_{2}$ (X = Ca, Sr, Ba, Yb). We attribute this difference to the antiferromagnetic ordering of Eu atoms, which modifies the topological properties despite the preservation of Kramers degeneracy in compounds adopting the CaAl$_2$Si$_2$-type structure.
However, when we include the hybrid calculations, EuCd$_2$As$_2$ becomes a trivial insulator.
Instead, EuCd$_2$Bi$_2$ exhibits a different behavior. Indeed, we have previously shown that EuCd$_2$Bi$_2$ exhibits a band inversion at the $\Gamma$ point and may therefore host nontrivial topological states~\cite{Cuono23EuCd2As2} even within hybrid functionals. Despite the larger spin-orbit coupling of Bi compared to As, the conduction band in this compound is significantly narrower, being derived primarily from the Mg$^{2+}$ $3s$-states rather than the Cd$^{2+}$ $5s$-states. This reduction in bandwidth compensates for the enhanced spin-orbit coupling associated with Bi. As a result, EuCd$_2$Bi$_2$ is one of the few Eu-based materials that has been proposed as a potential topological material~\cite{Cuono23EuCd2As2}.

\begin{figure}
    \centering
    \includegraphics[width=1\linewidth]{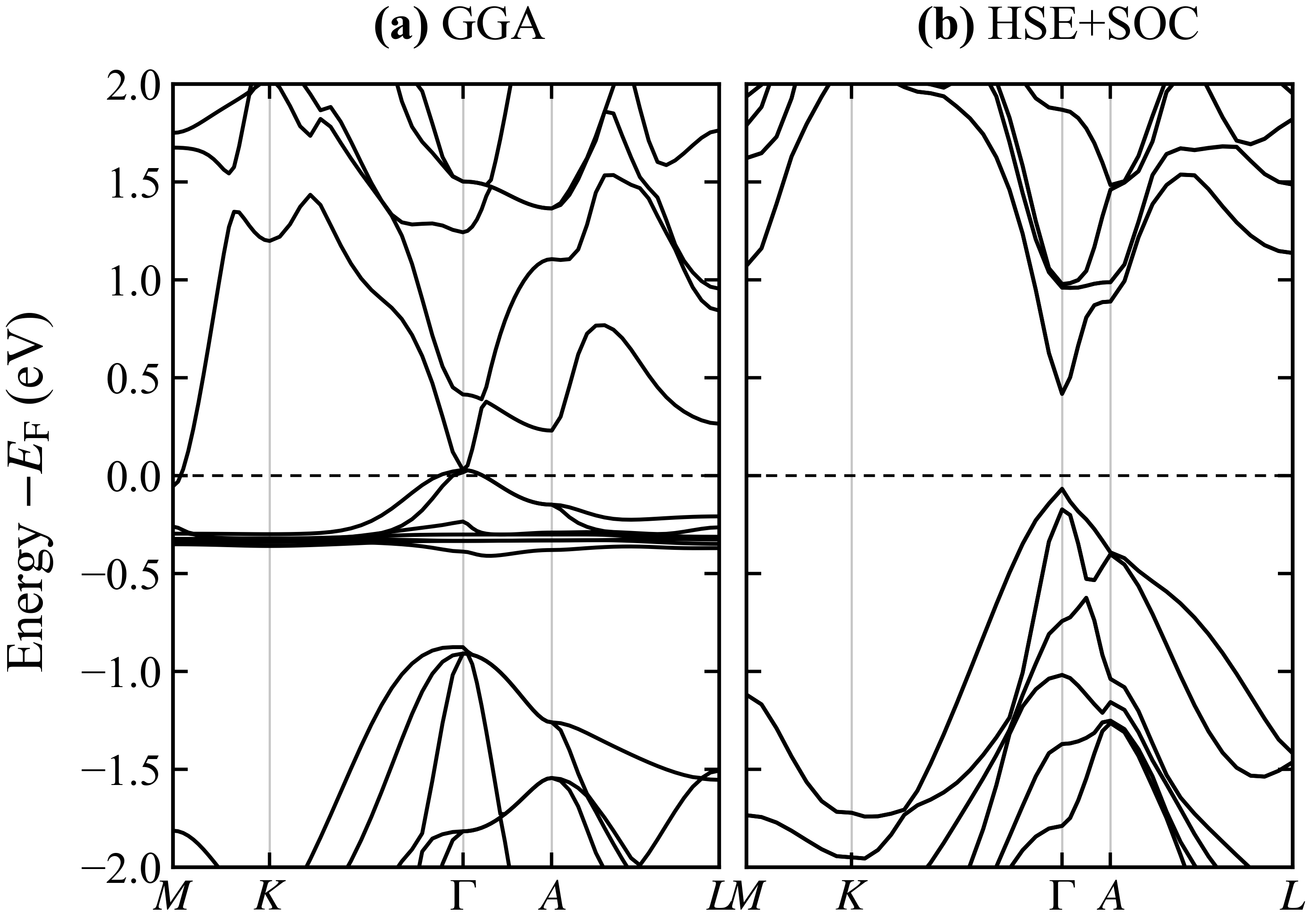}
    \caption{Relativistic band structures within (a) GGA and (b) HSE of EuMg$_2$Bi$_2$ along the high-symmetry path M--K--$\Gamma$--A--L. The Néel vector is along the y-direction. The energy range spans from $-2$ to $+2$~eV, with the Fermi level set to zero, indicated by a dashed line.}
    \label{fig:Figure5}
\end{figure}

We present the band structure of EuMg$_2$Bi$_2$ within the GGA framework in Fig. \ref{fig:Figure5}(a) and within the HSE exchange-correlation functional in Fig. \ref{fig:Figure5}(b). In GGA, we observe several flat bands between $-0.3$ and $-0.4$ eV, corresponding to the Eu $4f$ bands. In the HSE calculations, these bands are shifted away from the Fermi level, in agreement with experimental observations.
The topological invariant $\mathbb{Z}_4$ obtained within GGA is $\mathbb{Z}_4=2$, which is analogous to the value calculated for EuCd$_2$As$_2$ within GGA+U and it represents the axion insulating phase\cite{Ma:2020_AM}. The GGA+U calculations for EuMg$_2$Bi$_2$  do not alter the topological invariant compared with GGA. In contrast, the system becomes topologically trivial with $\mathbb{Z}_2=\mathbb{Z}_4=0$ when treated using the HSE exchange-correlation functional.
Since HSE has successfully reproduced the experimental results for other Eu compounds\cite{Cuono23EuCd2As2,PhysRevLett.131.186704}, we expect that it will also provide a more accurate description of the experimental situation in this case.
Within HSE, EuMg$_2$Bi$_2$ is a trivial insulator with a direct band gap of 0.24 eV at $\Gamma$ point.
The calculated band gap is robust against different orientations of the Néel vector; these results are not shown because the differences are difficult to discern. Therefore, EuMg$_2$Bi$_2$ remains a narrow-gap trivial insulator regardless of the Néel vector orientation. By combining the present results with those reported in the literature~\cite{Cuono23EuCd2As2}, we find that EuCd$_2$As$_2$ and EuMg$_2$Bi$_2$ are topologically trivial, whereas EuZn$_2$Bi$_2$ and EuCd$_2$Bi$_2$ are topologically nontrivial. Although using Bi as the anion is a promising direction, combining it with a light element such as Mg, whose 3s orbitals contribute to the conduction bands, does not result in the desired topological character. Instead, Bi should be combined with heavier elements such as Zn or Cd, whose 4s and 5s orbitals, respectively, contribute to the conduction bands, as in EuZn$_2$Bi$_2$ and EuCd$_2$Bi$_2$.

\section{Effect of strain on topological properties}

From the band structure and the calculation of the topological invariant, we conclude that the only time-reversal-invariant point at which a band inversion may occur is the $\Gamma$ point. Therefore, to evaluate the topological properties, it is sufficient to search for a closing of the direct band gap at the $\Gamma$ point. To induce a topological phase transition, we apply uniaxial strain and hydrostatic pressure in this section.
We define the uniaxial strain along the $c$ axis, $\varepsilon_c$, as
$\varepsilon_c = \frac{c - c_0}{c_0}$, where $c_0$ is the lattice parameter of the unstrained structure.
We define the hydrostatic (volumetric) strain, $\varepsilon_V$, as
$\varepsilon_V = \frac{V - V_0}{V_0}$, where $V_0$ is the volume of the unstrained structure. Therefore, positive values of $\varepsilon_c$ and $\varepsilon_V$ correspond to tensile strain, whereas negative values correspond to compressive strain. In the last subsection, we calculate the band-gap deformation potential.

\begin{figure}
    \centering
    \includegraphics[width=1\linewidth]{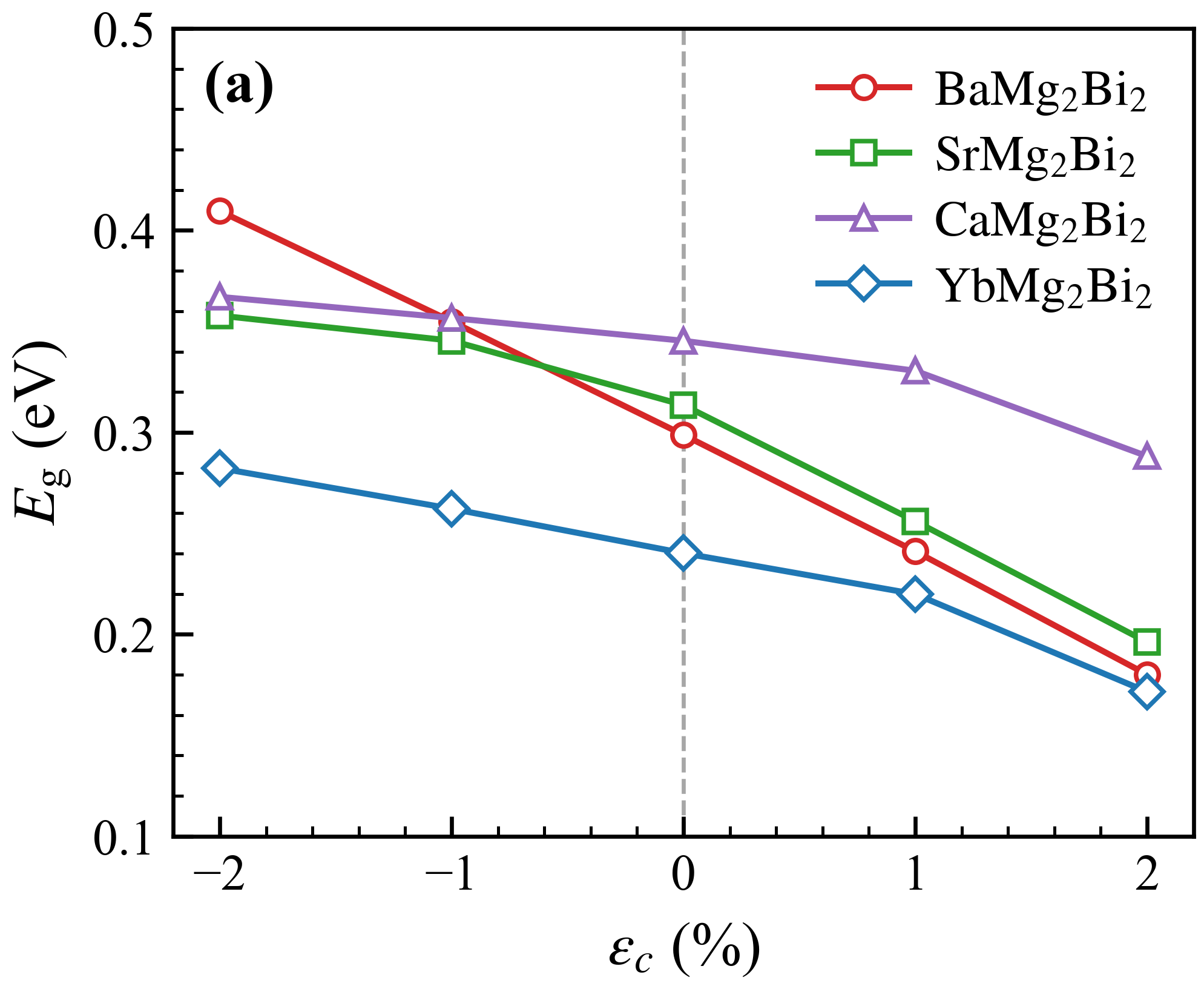}
    \includegraphics[width=1\linewidth]{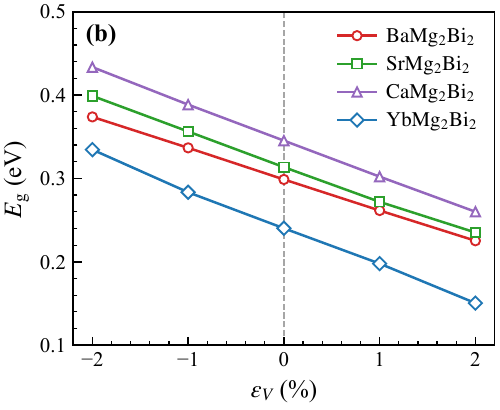}
    \caption{(a) Energy gap $E_g$ as a function of the uniaxial strain along the $c$ axis, $\varepsilon_c$, for the selected $\mathrm{XMg_2Bi_2}$ compounds ($X = \mathrm{Ca}$, $\mathrm{Sr}$, $\mathrm{Ba}$, and $\mathrm{Yb}$). (b) Energy gap $E_g$ as a function of the uniform hydrostatic pressure, $\varepsilon_V$, for the selected $\mathrm{XMg_2Bi_2}$ compounds ($X = \mathrm{Ca}$, $\mathrm{Sr}$, $\mathrm{Ba}$, and $\mathrm{Yb}$).}
    \label{fig:gap_cstrain_PRB}
\end{figure}

\subsection{Bulk under strain along the c-axis}

We calculate the direct band gap at $\Gamma$ and the topological invariant as a function of the compressive strain along the c-axis for XMg$_{2}$Bi$_{2}$ (X = Ca, Sr, Ba, Yb). The results are reported in Fig. \ref{fig:gap_cstrain_PRB}(a).
We observe a general trend in which the trivial band gap decreases linearly with compressive/tensile strain. 

At zero strain, CaMg$_2$Bi$_2$ exhibits the largest trivial band gap among the investigated compounds, with a value of approximately 0.35~eV, whereas YbMg$_2$Bi$_2$ presents the smallest band gap, of about 0.25~eV. The direct trivial band gap decreases as the atomic weight of the X element increases. This relatively small variation suggests that the electronic structure of these compounds is only weakly affected by the different cation species. This behavior can be attributed to the orbital character of the band edges, where the conduction band minimum is mainly derived from the 3$s$-states of Mg, while the valence band maximum originates primarily from the 6$p$-states of Bi. Consequently, the substitution of the cation has a limited impact on the fundamental band-gap size.

Extrapolating the linear behavior of the gap reported in Fig. \ref{fig:gap_cstrain_PRB}(a) for BaMg$_2$Bi$_2$, we obtain that the band gap closure and the emergence of the topological phase should appear for tensile strains superior to 6\%, which are not feasible to achieve in materials.

\begin{figure}    
    \includegraphics[width=0.99\linewidth]{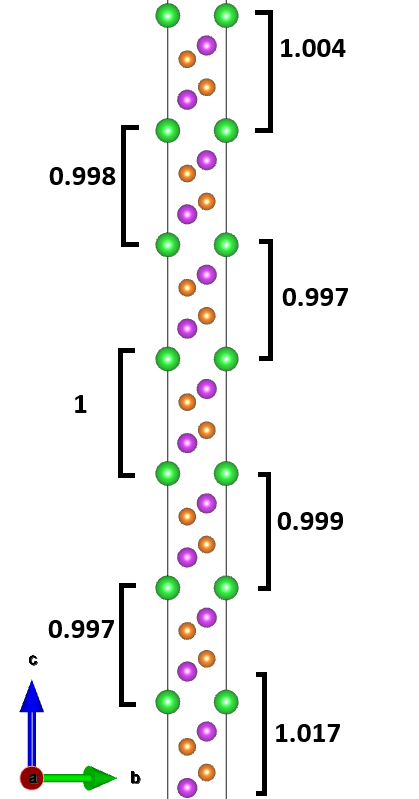}
    \caption{Asymmetric slab consisting of 7 unit cells of BaMg$_2$Bi$_2$ along the (001) direction, constrained in the $ab$ plane and relaxed along the $c$ direction. The top surface is Ba-terminated, whereas the bottom surface is Bi-terminated. The effective out-of-plane lattice constant of the inner layer was normalized to 1, while the ratios of the other out-of-plane lattice constants are reported in the figure. }
    \label{fig:Figure7}
\end{figure}

Slab calculations were performed by fixing the in-plane lattice constant while allowing relaxation of the atomic $z$ coordinates, and consequently of the effective out-of-plane lattice constant, for each layer. Owing to the large size of the system, the use of hybrid functionals is computationally prohibitive; however, the standard GGA functional is sufficient to accurately describe the structural properties.
We then compare the out-of-plane lattice constant of the inner layer, $c_{\mathrm{inn}}$, with those of the surface layers. As shown in Fig. \ref{fig:Figure7}, the subsurface layers exhibit out-of-plane lattice constants ranging from $0.997c_{\mathrm{inn}}$ to $0.999c_{\mathrm{inn}}$, which drives the system away from the topological phase. The out-of-plane lattice constant of the Ba-terminated surface is $c_{\mathrm{Ba-sur}}=1.004c_{\mathrm{inn}}$, while that of the Bi-terminated surface is $c_{\mathrm{Bi-sur}}=1.017c_{\mathrm{inn}}$.
The optimized slabs exhibit outward surface relaxation, manifested as an expansion of the interlayer spacing normal to the surface. This outward relaxation results in a positive $\varepsilon_c$, but its magnitude is too small to induce a topological transition. Furthermore, previous studies have shown that films only a few nanometers thick may not retain the topological properties of the bulk system, as the surface layers can become topologically trivial due to bandwidth reduction.\cite{Bernevig:2006_S,PhysRevB.107.045138}

Moreover, on the surface of insulators, the effective Coulomb repulsion increases due to the combined effects of bandwidth narrowing and reduced screening, which enhance electron--electron interactions\cite{PhysRevLett.109.146401} and favor the opening of a trivial band gap.
Therefore, we conclude that surface effects cannot drive these systems into a topological phase.

\subsection{Bulk under hydrostatic pressure}

We calculate the direct band gap at $\Gamma$ and the topological invariant as a function of the hydrostatic pressure for XMg$_{2}$Bi$_{2}$ (X = Ca, Sr, Ba, Yb). The results are reported in Fig. \ref{fig:gap_cstrain_PRB}(b).

The behaviour of the band gap under hydrostatic pressure is more uniform among the compounds with respect to the uniaxial strain, exhibiting a similar slope as a function of volume. By extrapolating the trend observed in the curve, we estimate that an increase in the unit-cell volume of approximately 4\% could induce a transition of YbMg$_{2}$Bi$_{2}$ into a topological phase. Although this value is obtained through extrapolation, it indicates that lattice expansions may significantly modify the electronic structure and promote band inversion. As a result, we conclude that neither uniaxial strain nor hydrostatic pressure is likely to induce a topological phase transition in the class of materials XMg$_{2}$Bi$_{2}$ (X = Ca, Sr, Ba, Yb) for realistic values of these perturbations.

\subsection{Band gap deformation potential}

To quantify the sensitivity of the electronic structure to lattice distortions, we characterize the strain dependence of the direct band gap through the band-gap deformation potential. Following the definition commonly adopted in first-principles studies of strain-induced band-gap variations \cite{PhysRevB.94.245411,PhysRevB.89.195135} we define the deformation potential as the derivative of the direct band gap with respect to the applied strain. For uniaxial strain along the crystallographic $c$ axis, the deformation potential is given by
\begin{equation}
D_c=\frac{\partial E_g}{\partial \varepsilon_c},
\end{equation}
where $E_g$ is the direct band gap at the $\Gamma$ point. Similarly, for hydrostatic (volumetric) strain, the corresponding deformation potential is defined as
\begin{equation}
D_V=\frac{\partial E_g}{\partial \varepsilon_V}.
\end{equation}
Negative values of $D_c$ or $D_V$ indicate that tensile strain reduces the band gap, whereas positive values correspond to an increase of the band gap under lattice expansion. In practice, the deformation potentials are obtained from the slope of the calculated band gap dependence on the corresponding strain. For the considered systems, all values of $D_c$ and $D_V$ are negative, as listed in Table~\ref{Deformation_potential}. The values of $D_V$ are relatively uniform, ranging from -3.8 to -4.3 for the selected compound. In contrast, the values of $D_c$ for the other compounds span a much wider range, from -1.3 to -5.7.

\begin{table}[htbp]
\centering
\renewcommand{\arraystretch}{1.2}
\caption{Band gap deformation potentials of non-magnetic XMg$_2$Bi$_2$
(X = Ba, Ca, Sr, and Yb) for uniaxial strain along the crystallographic $c$ axis $D_c$ and for hydrostatic strain $D_V$.}
\label{Deformation_potential}
\setlength{\tabcolsep}{8pt}
\begin{tabular}{lcc}
\hline
Compound & $D_c$ (eV) & $D_V$ (eV) \\
\hline
BaMg$_2$Bi$_2$ & $-5.7$ & $-3.8$ \\
CaMg$_2$Bi$_2$ & $-1.3$ & $-4.3$ \\
SrMg$_2$Bi$_2$ & $-4.5$ & $-4.2$ \\
YbMg$_2$Bi$_2$ & $-2.1$ & $-4.3$ \\
%EuMg$_2$Bi$_2$ & $- $ & $- $ \\
\hline
\end{tabular}
\end{table}

\vspace{1cm}
\section{Discussion and Conclusions}
We have revisited the topological properties of XMg$_2$Bi$_2$ (X=Ca, Sr, Ba, Yb and Eu) by computing the topological invariants within GGA and within hybrid functional calculations.
Within the less accurate GGA approximation, the nonmagnetic systems are predicted to be three-dimensional topological Dirac semimetals, while the antiferromagnetic EuMg$_2$Bi$_2$ additionally hosts an axion insulating phase. However, experimental observations are expected to be more consistent with the more accurate hybrid-functional calculations. 
We demonstrate that bulk XMg$_2$Bi$_2$ (X=Ca, Sr, Ba, Yb and Eu) with a CaAl$_2$Si$_2$-type crystal structure are trivial insulators with a direct band gap between 0.25 and 0.35 eV at the $\Gamma$ point within hybrid functionals. 
The direct trivial band gap exhibits a decreasing trend as the atomic weight of the X element increases.
For the antiferromagnetic compound with Eu, these results are robust with respect to the N\'eel vector.
The same trivial narrow-gap semiconducting phases are obtained for these materials under surface relaxation, uniaxial strain, and hydrostatic pressure, where large, unrealistic values of strain and pressure are required to induce the transition to a topologically non-trivial phase. Beyond the electronic band structure and band inversion, our results are corroborated by the calculation of the topological invariant. Therefore, we conclude that this class of materials represents robust narrow-gap semiconductors.

Our results demonstrate that the DFT-predicted topological behavior in XMg$_2$Bi$_2$ (X = Ca, Sr, Ba, Yb and Eu) with the CaAl$_2$Si$_2$-type crystal structure is an artifact arising from the inadequate treatment of electronic correlations within standard density functional theory. In contrast, hybrid functional calculations more accurately account for the stronger electronic correlations, leading to a trivial insulating phase. If we consider our previous results on EuCd$_2$As$_2$ and other materials\cite{Cuono23EuCd2As2}, we can conclude that semilocal DFT systematically overestimates inversion in CaAl$_2$Si$_2$ compounds.

Even the observation of surface states is not enough to claim a topological behaviour, because the system can host trivial surface states without a Dirac point and without a non-trivial $\mathbb{Z}_2$ topological invariant.
Eventual topology revealed experimentally in this class of materials cannot be attributed to intrinsic band structure. Still, it should be attributed to extrinsic effects, such as doping or surface reconstructions. These effects could change the energy levels and induce topology, but the stoichiometric compounds XMg$_2$Bi$_2$ with an ideal CaAl$_2$Si$_2$-type crystal structure are topologically trivial. 

\vspace{1cm}
\section*{Acknowledgments}
This research was supported by the "MagTop" project (FENG.02.01-IP.05-0028/23) carried out within the "International Research
Agendas" programme of the Foundation for Polish Science, co-financed by the European Union under the European Funds for Smart Economy 2021-2027 (FENG). 
A. Facca and A. S. were supported by the National Science Centre (NCN), Poland, under grant OPUS 21 No. UMO-2021/41/B/ST3/04475.
The author further acknowledges access to the computing facilities of the Interdisciplinary Center of Modeling at the University of Warsaw, Grant g91-1418, g91-1419, g96-1808, g96-1809, g103-2540, g104-2571, g104-2572 and g104-2573 for the availability of high-performance computing resources and support. We acknowledge the access to the computing facilities of the Poznan Supercomputing and Networking Center, Grants No. pl0267-01, pl0365-01, pl0471-01 and pl0694-01.\\

\appendix

\section{Computational details}

\begin{table}[htbp]
\centering
\renewcommand{\arraystretch}{1.8}
\caption{Optimized structural parameters of XMg$_2$Bi$_2$
(X = Ba, Ca, Sr, Yb, and Eu). All compounds crystallize in the
$P\bar{3}m1$ space group (No.~164). The atoms occupy the Wyckoff
positions X ($1a$), Mg ($2d$), and Bi ($2d$), with fractional coordinates
X: $(0,0,0)$, Mg: $(\frac{2}{3},\frac{1}{3},z_{\rm Mg})$, and
Bi: $(\frac{1}{3},\frac{2}{3},z_{\rm Bi})$.}
\label{table_lattice_BaMg2Bi2}
\setlength{\tabcolsep}{5.5pt} 
\begin{tabular}{lccccc}
\hline
Compound & $a$ (\AA) & $c$ (\AA) & $c/a$ & $z_{\rm Mg}$ & $z_{\rm Bi}$ \\
\hline
BaMg$_2$Bi$_2$ & 4.897 & 8.316 & 1.698 & 0.377772 & 0.265281\\
CaMg$_2$Bi$_2$ & 4.768 & 7.731 & 1.621 & 0.372270 & 0.241644\\
SrMg$_2$Bi$_2$ & 4.831 & 7.998 & 1.655 & 0.374857 & 0.253211\\
YbMg$_2$Bi$_2$ & 4.762 & 7.697 & 1.616 & 0.371126 & 0.239526\\
EuMg$_2$Bi$_2$ & 4.849 & 7.891 & 1.627 & 0.369527 & 0.248979\\
\hline
\end{tabular}
\end{table}

First-principles calculations were performed using the Vienna Ab~initio Simulation Package (VASP)~\cite{kresse1993ab,kresse1996efficiency}, 
within the framework of density functional theory and employing the projector augmented-wave method~\cite{kresse1999ultrasoft}. 
The exchange--correlation potential was described using the generalized gradient approximation with the Perdew--Burke--Ernzerhof (PBE) functional~\cite{Perdew96}. 
A plane-wave cutoff energy of 300~eV was used throughout, and the total energy convergence criterion was set to $10^{-6}$~eV. 
First-principles calculations were carried out using the HSE06 hybrid functional\cite{10.1063/1.2404663}, which incorporates a screened fraction of exact exchange into the PBE generalized gradient approximation. The standard screening parameter and mixing fraction were employed. The k-grid used was $6\times6\times4$ for the bulk, while it was $6\times6\times4$ for the slab. Since the Fermi surface is absent in the trivial insulating phase and composed of the two Dirac points in the three-dimensional topological semimetallic phase, increasing the number of k-points does not lead to significant changes in the system's properties. For visualization purposes, the Mg-$s$ orbital weight was scaled by a factor of 30 in the band structure of BaMg$_2$Bi$_2$ in Fig. \ref{fig:Figure4}.

The compound crystallizes in the CaAl$_2$Si$_2$-type structure with the centrosymmetric space group P$\bar{3}$m1 (No. 164). The corresponding Brillouin zone contains four time-reversal invariant momenta: one $\Gamma$ point, three symmetry-equivalent M points, one A point, and three symmetry-equivalent L points. Owing to the crystal symmetry, all M points are symmetry equivalent, as are all L points; therefore, it is sufficient to consider a single representative M point and a single representative L point when evaluating the $\mathbb{Z}_2$ topological invariant in the presence of inversion symmetry. The topological invariant was calculated using the IrRep package\cite{IRAOLA2022108226}. The lattice constants and the Wyckoff positions for the selected compounds were reported in Table \ref{table_lattice_BaMg2Bi2}.

\bibliography{references}
\end{document}